\documentclass{article}
\usepackage{spconf,amsmath,graphicx,xcolor}
\usepackage{hyperref}
\usepackage{enumitem}
\usepackage{multirow}

\title{A Meta-GNN approach to personalized seizure detection and classification}
\name{Abdellah Rahmani\space\space\space\space\space\space\space Arun Venkitaraman\space\space\space\space\space\space\space Pascal Frossard}
\address{Signal Processing Laboratory LTS4, EPFL, Lausanne, Switzerland}
\begin{document}
%
\maketitle
\begin{abstract}
In this paper, we propose a personalized seizure detection and classification framework that quickly adapts to a specific patient from limited seizure samples. We achieve this by combining two novel paradigms that have recently seen much success in a wide variety of real-world applications: graph neural networks (GNN), and meta-learning. We train a Meta-GNN based classifier that learns a global model from a set of training patients such that this global model can eventually be adapted to a new unseen patient using very limited samples. 
We apply our approach on the TUSZ-dataset, one of the largest and publicly available benchmark datasets for epilepsy. We show that our method outperforms the baselines by reaching 82.7\% on accuracy and 82.08\% on F1 score after only 20 iterations on new unseen patients. 

\end{abstract}
\section{Introduction}
\label{sec:intro}

Epilepsy affects between 60 and 70 millions people worldwide and about 3 millions are diagnosed with this disease every year \cite{alarcon2012introduction,delanty1998medical,sazgar2019absolute}. It is estimated that up to 70\% of people suffering from epilepsy could control their illness if they are properly diagnosed and treated with anti-epileptic drugs \cite{Who}. In the rest of the cases where the patients do not respond to drugs, constant monitoring is crucial. The monitoring is usually done through measurements of brain signals using electroencephalograms (EEG) that record the electrical signals produced by different brain regions through electrodes placed over the head. Monitoring is done manually using the knowledge and experience of experts based on visual inspection of the EEG recordings. Given that this process tend to be extremely resource and time-intensive, researchers are increasingly interested in the development of automated approaches.
It has been shown recently that deep learning based models can help achieve fairly impressive performances in seizure detection \cite{rasheed2020machine,saab2020weak,siddiqui2020review, o2020neonatal} and classification tasks \cite{ahmedt2020neural,raghu2020eeg,roy2020seizure,ievsmantas2020convolutional}.

A vast majority of these previous works use deep learning systems in the form of Convolutional Neural Networks (CNN) to extract relevant features from the EEG signals and to perform the detection or classification tasks. However, CNNs are usually agnostic to the inherent geometry of the multi-channel brain signals and do not actively exploit any structural information in the data. In particular, the different EEG electrodes share a lot of information functionally or structurally due to their close positions on the head. It is known that in such cases where there is inherent geometry involved, graph-based convolutions or graph neural networks can supersede CNNs in performance \cite{kipf2016semi}. Recently,  Tang et al. \cite{tang2021self} proposed a self-supervised approach to seizure detection in the form of graph neural networks that actively use the underlying graph structure of the EEG channels  networks. By applying their approach to Temple University Hospital EEG Seizure Corpus (TUSZ) \cite{obeid2016temple, shah2018temple}, one of the largest public datasets on adult seizures known to have a lot of diversity in terms of seizures and patient types, they showed that graph-based modelling approaches can significantly improve the performance of seizure detection and classification models. Nevertheless, this approach learns one single model for all patients, which may cause in some cases a degradation on new unseen patients, due to individual differences. Moreover, the approach is not directly applicable to settings where we have to learn from limited seizure data, which is closer to the realistic settings in hospitals.

Since our motivation is to develop a personalization approach, we use relatively simpler GNNs for our experiments unlike the ones used in Tang et al. In particular, we use two of the most popular GNN architectures: graph convolution networks (GCN) \cite{kipf2016semi} and graph attention networks (GAT) \cite{velivckovic2017graph}. Unlike the GCNs, which inherently work with the fixed initial graph structure, GATs extract a new functional graph for every data sample using attention. 
We personalize the GNN to the different patients using the popular meta-learning approach called MAML (Model-Agnostic Meta-Learning \cite{finn2017model}) that extracts a single global model from a set of training patients which is then fine-tuned and personalized to new patients using very limited new samples.  We propose a new updated loss function in order to overcome the  overfitting issues encountered in the original MAML. We apply our approach to both seizure detection (a binary classification of non seizures versus seizures) and classification (between three classes of focal seizure, generalized seizure, and non-seizure) tasks. We evaluate our model on TUSZ \cite{obeid2016temple, shah2018temple}, and show that our model outperforms both a single GNN model per patient, and one single GNN model for all the patients and achieves comparable results to Saab et al. \cite{saab2020weak} model (one of the state-of-the-art model for seizure detection) but in less training time. In other words, we show that using one Meta-GNN that adapts to patients offers great potential.

\section{Preliminaries}

\label{sec:format}
\subsection{Modelling EEGs as a graph signal}
\label{sec:graph}
Following the modelling in a previous study \cite{tang2021self}, we represent a segment of EEG (that we later refer to as 'clip') as a signal on a graph  $\mathcal{G}=\{\mathcal{V}, \mathcal{E}, \boldsymbol{W}\}$, where $\mathcal{V}$ is the set of vertices (i.e. EEG
electrodes/channels), $\mathcal{E}$ the set of edges and $\boldsymbol{W}$ the graph adjacency matrix. We compute the edge weights
$W_{ij}$ by applying a thresholded Gaussian kernel to the pairwise Euclidean
distance between the nodes (electrodes) $v_i$ and $v_j$ according to the standard 10-20 EEG electrode placement \cite{jasper1958ten}; $\boldsymbol{W}_{i j}=\exp \left(-\frac{\operatorname{dist}\left(v_i, v_j\right)^2}{\sigma^2}\right)$ if $\operatorname{dist}\left(v_i, v_j\right) \leq \kappa$, else 0. The parameter $\sigma$ is the standard deviation of the distances, and $\kappa$ is a threshold for sparsity. We choose the same threshold as \cite{tang2021self}, $\kappa = 0.9$. Each node (electrodes) incorporates a feature vector composed by Fourier coefficients of the corresponding recorded signal.
\subsection{Graph neural networks}
GNNs are well known architectures that deal with data which has non-trivial topology like graphs. Different deep learning techniques are extended to graph structured data such as graph convolution (GCN) \cite{kipf2016semi} and also graph attention (GAT) \cite{velivckovic2017graph}.

\textbf{Graph Convolution Networks (GCN):} there are different types of GCNs. In our architecture we focused on GCN introduced by Kipf et al. \cite{kipf2016semi}. For this model, the goal is to learn a function of features on a graph $\mathcal{G}$,  which takes as input: 

- $\mathbf{X}$ : which is a node feature matrix $(N,F)$, where every raw represents a node feature.

-  $\mathbf{A}$ : The adjacency matrix of the graph $\mathcal{G}$. 

and outputs 
$
 \mathbf{X}^{\prime}=\hat{\mathbf{D}}^{-1 / 2} \hat{\mathbf{A}} \hat{\mathbf{D}}^{-1 / 2} \mathbf{X} \boldsymbol{\Theta}
$, where $\hat{\mathbf{A}}=\mathbf{A}+\mathbf{I}$ denotes the adjacency matrix with inserted self-loops and $\hat{D}_{i i}=\sum_{j=0} \hat{A}_{i j}$ its diagonal degree matrix.

\textbf{Graph Attention Networks (GAT):} these are neural network architectures that operate on graph-structured data, leveraging masked self-attentional layers \cite{vaswani2017attention}. The GAT output $\mathbf{x}_i^{\prime}$ for a target node $\mathbf{x}_i$ can be described as follows:
$
\mathbf{x}_i^{\prime}=\alpha_{i, i} \boldsymbol{\Theta} \mathbf{x}_i+\sum_{j \in \mathcal{N}(i)} \alpha_{i, j} \boldsymbol{\Theta} \mathbf{x}_j,
$
where $\boldsymbol{\Theta}$ is a learnable matrix of the model that is used to  transform the input features into higher-level features, $\mathcal{N}(i) = \{j \in \mathcal{V} | e_{ij} \neq 0\}$ is the neighbourhood of the node $i$ and the $\alpha_{i, j}$ are the attention coefficients. The attention mechanism in GAT is a single-layer feedforward neural network parametrized by a vector $\mathbf{a}$  as follows:\\
$
\alpha_{i, j}=\frac{\exp \left(\operatorname{LeakyReLU}\left(\mathbf{a}^{\top}\left[\boldsymbol{\Theta} \mathbf{x}_i \| \boldsymbol{\Theta} \mathbf{x}_j\right]\right)\right)}{\sum_{k \in \mathcal{N}(i) \cup\{i\}} \exp \left(\operatorname{LeakyReLU}\left(\mathbf{a}^{\top}\left[\boldsymbol{\Theta} \mathbf{x}_i \| \boldsymbol{\Theta} \mathbf{x}_k\right]\right)\right)},
$
where  LeakyReLU refers to the Leaky Rectified Linear activation \cite{maas2013rectifier} that is a learnable variant of the regular ReLU. 
\subsection{Meta learning}
We used the MAML model \cite{finn2017model}. We consider every patient as a different task, that is, we consider learning a seizure detector/classifier for a given patient from their data as the {\em task}. We train our architecture by using the MAML training procedure that uses a set of training tasks or patients.
The meta learning approach learns a global model that can rapidly adapt to new tasks using only few iterations. One of the popular meta-learning models is MAML \cite{finn2017model} and its particularity is that it can learn the parameters of any model in order to prepare it for fast adaptation. The models are optimized using gradient descent algorithm, and leverage this gradient-based learning rule to make a rapid
progress on new tasks drawn from the distribution over tasks. Mathematically, we consider a global model $f_\theta$ parameterized by $\theta$, the MAML training process proceeds by two steps:

\textbf{1. Adaptation (task-specific update):} 
For every task $\mathcal{T}_i$ drawn from a set of tasks $\Gamma$, we sample two data sets;
(a) the support set (training data)
$\mathcal{D}_s = \{(x_{s}^{1},y_{s}^{1}),...,(x_{s}^{n_s},y_{s}^{n_s})\}$ 
and 
(b) the query set (validation data) $\mathcal{D}_q= \{(x_{q}^{1},y_{q}^{1}),...,(x_{q}^{n_q}$ $,y_{q}^{n_q})\}$ where $n_s$ and $n_q$ are the numbers of data points in the support and query sets, and $(x,y)$ denote an input-output pair. In order to adapt the global model $f_\theta$ to the task $\mathcal{T}_i$, MAML proceeds by one or multiple gradient descent steps by using only the $\mathcal{D}_s$ as follows:
\begin{equation}
  \theta_i^{\prime}=\theta-\alpha \nabla_\theta \mathcal{L}(\theta,\mathcal{D}_s(\mathcal{T}_i)) ,
\end{equation}
where $\alpha$ is the inner learning rate, which is an hyper parameter that should be defined beforehand and $\mathcal{L}$ is the chosen loss for our problem.

\textbf{2. Meta optimization (Global update):}
After deriving $\theta_i^{\prime}$ for each task, we compute the gradients of the $f_{\theta_i^{\prime}}$ models using the data $\mathcal{D}_q$ and optimize $\theta$ as follows:
\begin{equation}
\label{eq2}
\theta = \theta-\beta \nabla_\theta \sum_{\mathcal{T}_i \in\Gamma} \mathcal{L}(\theta'_i,\mathcal{D}_q(\mathcal{T}_i))   , 
\end{equation}
where $\beta$ is the meta steps size. Using one or few gradient descent steps on $\theta$ produces $\theta_i^{\prime}$ which is specific and efficient for the task $i$. MAML training phase involves multiple iterations of performing both adaptation and meta-optimization that is obtained using a set of training tasks, training to convergence.

\section{Meta-GNN model}
\begin{figure}[t]
    \centering
    \includegraphics[scale=0.13]{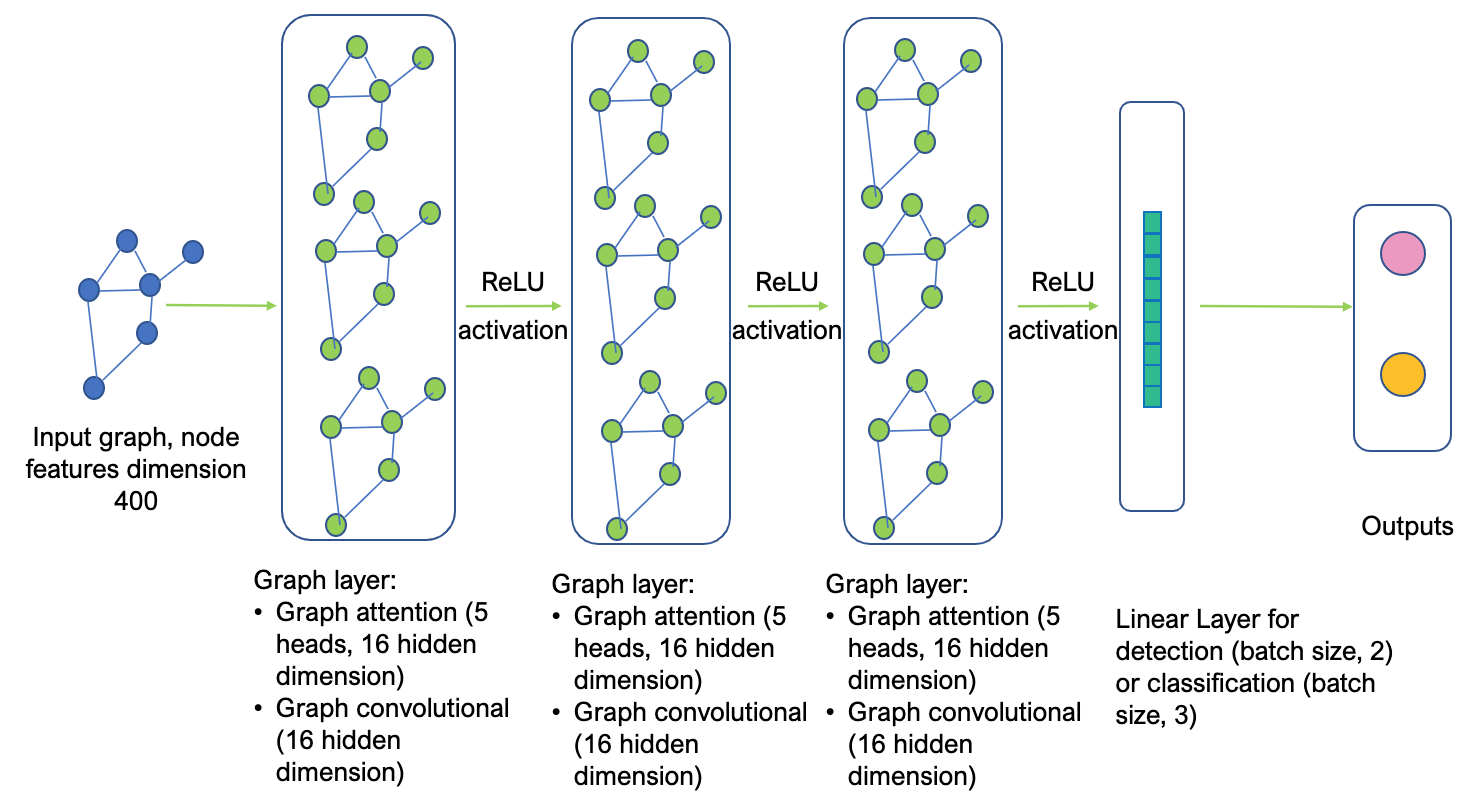}
    \caption{Proposed Graph neural network architecture.}
    \label{fig2}
\end{figure}
We propose to combine meta-learning ideas and GNNs to solve the detection and classification tasks. Our model takes the channel-wise features $\mathbf{X}$ for a given EEG segment or clip, and the distance graph $\mathcal{G}$ as input, and predicts the label  $\hat{y} = f_{\theta}(\mathbf{X},\mathcal{G})$ for the segment. We then use MAML \cite{finn2017model} described in Section 2 for personalization of the prediction to the different patients. In our case, every patient corresponds to a different task, and we perform meta-learning using a set of training patients. The global model $f_\theta$ is thus learnt from meta-training and eventually personalized using few samples of the patient's data to obtain the personalized or patient-specific model $f_\theta'$. The performance is then evaluated on a set of new unseen test patients.



However, it is known that meta learning model, particularly MAML, suffers from one of two types of over-fitting: 

(a) {\em  Memorization over-fitting}, in which the model is able to predict well the query set without relying on the support set but fails to generalize to new unseen task, and

(b) {\em Learner over-fitting}, in which the learner over-fits to the training set and does not generalize to the test set.

In our case, the patients represent non-mutually-exclusive tasks \cite{rajendran2020meta}, which is confirmed by the the performance of previous works that treated all the patient as a single task \cite{rasheed2020machine,saab2020weak,siddiqui2020review, o2020neonatal}. This mutual exclusivity causes a problem of memorization over-fitting. In our experiments, we noticed that the memorization overfitting strongly affect the test performance. In order to reduce memorization over-fitting, we propose the following modified MAML outer loop (Eq. \ref{eq2}):
\begin{equation}
\theta \leftarrow \theta-\beta \nabla_\theta \Big\{\sum_{\mathcal{T}_i \in \Gamma} \mathcal{L}(\theta'_i, \mathcal{D}_q(\mathcal{T}_i))+\gamma \mathcal{L}(\theta'_i,\mathcal{D}_s(\mathcal{T}_i))\Big\},
\end{equation}
where $\gamma$ is an hyper-parameter that controls the amount of information that we want to keep from the support set gradient directions. We want our model to succeed to classify well the query set but while relying also on the support set. The objective is to reduce the over-fitting on the query set of the training patients without impacting the performances on the query set. Minimizing the over-fitting will also enhance the performances of our model on new unseen patients.

\section{Experiments}
\label{sec:pagestyle}
\subsection{Dataset and Statistics}
%
%
%
%
%

Following the work in \cite{tang2021self}, we use the public Temple University Hospital EEG Seizure Corpus (TUSZ).  We include 19 EEG channels in the standard 10-20 system \cite{jasper1958ten}.
To train our architecture, we filter TUSZ dataset and we keep only patients with more than 4 seizures, but for testing we reduce the number to only 2 seizures. 
After this filtering process, we have 97 patients for training our model and 26 new patients (unseen) for testing. 

\textbf{Data preprocessing.\label{datapre}} We split every EEG recording to 10 seconds clips (windows) and compute their fast Fourier transform. Though potentially the size of the window also has a role to play, we restrict our analysis to 10 s windows here, and shall be reporting the results of the experiments on different window-sizes in the journal version of the work. In order to remove the artifacts, we perform a hard thresholding by keeping only the coefficients smaller than 40Hz and dropping all the others. Finally, we normalize these coefficients for every clip, by removing the mean over all channel and dividing by the standard deviation of the same clip, and use them as node features. 

\textbf{Training and testing procedure.} During the training phase, we consider all the available recordings from the training patients. During test time, we use limited data for the unseen test patients for fine-tuning (1 seizure of 1 minute duration which will be divided into 6 clips of 10 seconds i.e 12 samples of 6 seizures and 6 backgrounds). This is to simulate a more realistic setting where we want to quickly adapt existing global model to new patient in an efficient manner reducing the amount of time the new patients need to be under monitoring. The fine-tuned models are then evaluated based on the performance on the remaining data (about 1600 clips per test patient).

\textbf{Tasks.}
For the detection problem we use all the seizures that verify the aforementioned filtering condition and we have only 2 classes, the first one belongs to seizures and the second one to backgrounds (non seizures). In the other hand, for the seizure classification problem, we keep only two kinds of seizures the focal and the generalized seizure because these two are the most frequent classes in our dataset. We also keep the backgrounds, hence our model tries to distinguish between 3 different classes. The numbers of clips per class (either in detection or classification) are imbalanced, so, we over-sample the seizure clips by repeating them $N$ times where:
$\displaystyle N = \frac{\text{Number of background clips}}{\text{Number of seizure clips}}$.

\textbf{Baselines.} In the literature, there is no available baseline on few-shots learning settings. To compare our models GAT-Meta Learning (GAT-ML) and GCN-Meta Learning (GCN-ML) with other baselines, these two models share the same architecture (Figure \ref{fig2} but with two different layers,  we choose to build and train the following models:

(1) {\em Global GAT model (Glob-GAT):} we train one GAT model using all the data available from training patients and test on test patients.

(2) {\em Dense-CNN from Saab et al. per patient (CNN-PPAT):} we train one single Dense-CNN model described in \cite{saab2020weak} for every test patient.

(3) {\em GCN per patient (GCN-PPAT):} we train one single GCN model for every test patient.

(4) {\em GAT per patient (GAT-PPAT):} we train one single GAT model for every test patient.

We compare our models with GAT-PPAT and GCN-PPAT, that are models with the same architecture as ours but without prior knowledge which clarify the contribution of meta-learning approach. The comparison with Glob-GAT aims to highlight the importance of personalized models over global models. We also compare the performance against CNN-PPAT which is a state-of-the-art non-graph model.
\subsection{Experimental results}
\begin{figure}[t]
    \centering
    \includegraphics[scale=0.35]{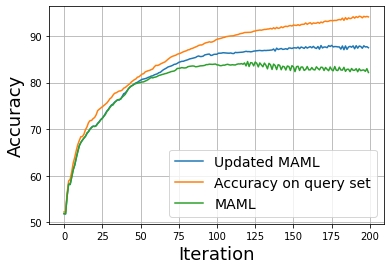}
    \caption{Training accuracy versus number of iterations. Blue line   - accuracy of the updated MAML on the support set;  while the Green -- accuracy of the original MAML on the same set; Orange -- accuracy of both model on query set.}
    \label{fig3}
    \vspace{-.15in}
\end{figure}
\begin{figure}
    \centering
    \includegraphics[scale=0.35]{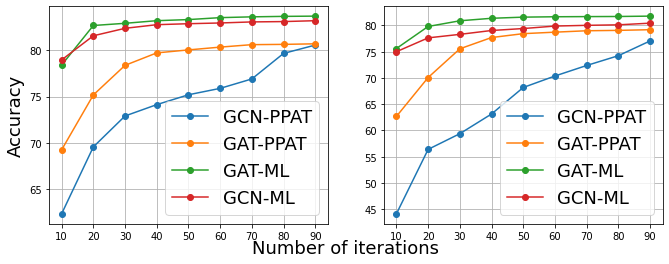}
    \caption{Average accuracy on test data with training iterations for detection (left) and classification (right).}
    \label{fig:model_iter}
\end{figure}

We compute all the metrics (Accuracy and Macro F1 score) we use in this paper via averaging the same metrics over the patients (either for training or testing). To see the effect of our training procedure, Figure \ref{fig3} shows that our updated version of MAML reduces the memorization overfitting by enhancing the accuracy on the support set across all training patients, without impacting the accuracy on the query set. Table \ref{tab:2} summarizes the results: First we can notice that for detection task, CNN-PPAT achieves 85.71\% (which is the best accuracy on this task), GAT-ML and GCN-ML  outperform the GAT-PPAT and GCN-PPAT. Notably, GAT in both meta learning framework or training single models without prior knowledge achieves better scores than GCN. The meta learning models (GAT-ML and GCN-ML) outperform GAT-PPAT with only 20 iterations while the GAT-PPAT is trained for 90 iterations with personalized early stopping (different early stopping for every patient). Figure \ref{fig:model_iter} shows the results for more fine tuning iterations, the average accuracy over the patients enhanced when we fine tune the meta learning models for more iterations and the difference between (GAT or GCN) -ML  and GAT-PPAT increases. GAT-ML also reaches 82.71\% with a small model and also in less training time compared to CNN-PPAT (training time: around 60 hours). Since CNN-PPAT is both a memory intensive
and computationally expensive, we report only the CNN-PPAT accuracy metric for detection in the current manuscript and shall report the remaining scores in a journal version. We notice the same behaviour for the F1 score metric. We do not plot it due to space limitation. For the classification task, the GAT-ML fine-tuned with 20 iterations outperforms the GAT-PPAT. However the 20 iterations were not enough for GCN-ML to outperform GAT-PPAT in a more complex setup than the detection setup. From the Figure \ref{fig:model_iter}, GCN-ML needs more than 50 iterations to outperform the GAT-PPAT. GAT tries to learn different edge weights for different clips which enhances the performance on the classification task. This shows that Assuming a static graph and relying only on node features to detect seizures seems to be feasible for classification, unlike for distinguishing between seizure types which is a more complicated task.
\begin{table}[t]
\begin{tabular}{l|c|c|c|c}
\hline Model (number  & \multicolumn{2}{|c|}{Detection} & \multicolumn{2}{c}{ Classification} \\
\cline { 2 - 5 } of iterations )& Acc.& F1 & Acc. & F1\\\hline
Glob-GAT (90) & $73.33$ & $72.00$ & $60.33$ & $43.54$ \\\hline 
CNN-PPAT (90)& $\mathbf{85.71}$& - & - & -\\  

GCN-PPAT (90) & $80.58$ & $79.88$ & $77.04$ & $76.16$ \\

GAT-PPAT (90) & $80.71$ & $80.51$ & $79.19$ & $78.54$ \\
\hline
GCN-ML (20) & $81.59$ & $81.36$ & $77.63$ & $76.49$ \\
 GAT-ML (20) & $\mathbf{82.70}$ & $\mathbf{82.08}$ & $\mathbf{79.80}$ & $\mathbf{79.28} $ \\
GCN-ML (90) & $83.22$ & $83.04$ & $80.43$ & $79.38$ \\
 GAT-ML (90) & $\mathbf{83.72}$ & $\mathbf{83.20}$ & $\mathbf{81.77}$ & $\mathbf{81.26} $ \\
\hline
\end{tabular}
    \caption{Comparison between the baselines and our model in detection and classification setting using accuracy (Acc.) and Macro F1 metrics on test data (unseen patients).}
    \label{tab:2}
\end{table}

\section{Conclusion}
We presented a method that uses graph neural networks combined with meta learning techniques for epilepsy detection and classification tasks. We showed how our approach achieves personalized diagnosis on new patients at test time using only a few new samples from the patient and within a few iterations. We showed that our  Meta-GNN model outperforms graph-based models that are trained per patient and also the single GNN model that is trained on all the patients. Furthermore, our experiments indicate that that personalized graphs learnt from the data are better than static (distance) based graphs for distinguishing between seizure types. In the future, we plan to study these aspects in more detail. 
\bibliographystyle{IEEEbib}
\bibliography{strings}

\begin{thebibliography}{10}

\bibitem{alarcon2012introduction}
Gonzalo Alarc{\'o}n and Antonio Valent{\'\i}n,
\newblock {\em Introduction to epilepsy},
\newblock Cambridge University Press, 2012.

\bibitem{delanty1998medical}
Norman Delanty, Carl~J Vaughan, and Jacqueline~A French,
\newblock ``Medical causes of seizures,''
\newblock {\em The Lancet}, vol. 352, no. 9125, pp. 383--390, 1998.

\bibitem{sazgar2019absolute}
Mona Sazgar and Michael~G Young,
\newblock {\em Absolute epilepsy and EEG rotation review: Essentials for
  trainees},
\newblock Springer, 2019.

\bibitem{Who}
``World health organization. epilepsy (2022),''
  \url{https://www.who.int/en/news-room/fact-sheets/detail/epilepsy}.

\bibitem{rasheed2020machine}
Khansa Rasheed, Adnan Qayyum, Junaid Qadir, and et~al.,
\newblock ``Machine learning for predicting epileptic seizures using eeg
  signals: A review,''
\newblock {\em IEEE Reviews in Biomedical Engineering}, vol. 14, pp. 139--155,
  2020.

\bibitem{saab2020weak}
Khaled Saab, Jared Dunnmon, Christopher R{\'e}, Daniel Rubin, and Christopher
  Lee-Messer,
\newblock ``Weak supervision as an efficient approach for automated seizure
  detection in electroencephalography,''
\newblock {\em NPJ digital medicine}, vol. 3, no. 1, pp. 1--12, 2020.

\bibitem{siddiqui2020review}
Mohammad~Khubeb Siddiqui, Ruben Morales-Menendez, Xiaodi Huang, and Nasir
  Hussain,
\newblock ``A review of epileptic seizure detection using machine learning
  classifiers,''
\newblock {\em Brain informatics}, vol. 7, no. 1, pp. 1--18, 2020.

\bibitem{o2020neonatal}
Alison O’Shea, Gordon Lightbody, Geraldine Boylan, and Andriy Temko,
\newblock ``Neonatal seizure detection from raw multi-channel eeg using a fully
  convolutional architecture,''
\newblock {\em Neural Networks}, vol. 123, pp. 12--25, 2020.

\bibitem{ahmedt2020neural}
David Ahmedt-Aristizabal, Tharindu Fernando, Simon Denman, Lars Petersson,
  Matthew~J Aburn, and Clinton Fookes,
\newblock ``Neural memory networks for seizure type classification,''
\newblock in {\em 2020 IEEE Engineering in Medicine \& Biology Society (EMBC)}.
  IEEE, 2020, pp. 569--575.

\bibitem{raghu2020eeg}
Shivarudhrappa Raghu, Natarajan Sriraam, and Yasin et~al. Temel,
\newblock ``{EEG} based multi-class seizure type classification using
  convolutional neural network and transfer learning,''
\newblock {\em Neural Networks}, vol. 124, pp. 202--212, 2020.

\bibitem{roy2020seizure}
Subhrajit Roy, Umar Asif, Jianbin Tang, and Stefan Harrer,
\newblock ``Seizure type classification using eeg signals and machine learning:
  Setting a benchmark,''
\newblock in {\em 2020 IEEE Signal Processing in Medicine and Biology Symposium
  (SPMB)}. IEEE, 2020, pp. 1--6.

\bibitem{ievsmantas2020convolutional}
Tomas Ie{\v{s}}mantas and Robertas Alzbutas,
\newblock ``Convolutional neural network for detection and classification of
  seizures in clinical data,''
\newblock {\em Medical \& Biological Engineering \& Computing}, vol. 58, no. 9,
  pp. 1919--1932, 2020.

\bibitem{kipf2016semi}
Thomas~N Kipf and Max Welling,
\newblock ``Semi-supervised classification with graph convolutional networks,''
\newblock {\em arXiv preprint arXiv:1609.02907}, 2016.

\bibitem{tang2021self}
Siyi Tang, Jared Dunnmon, and Khaled Kamal et~al. Saab,
\newblock ``Self-supervised graph neural networks for improved
  electroencephalographic seizure analysis,''
\newblock in {\em International Conference on Learning Representations}, 2021.

\bibitem{obeid2016temple}
Iyad Obeid and Joseph Picone,
\newblock ``The temple university hospital eeg data corpus,''
\newblock {\em Frontiers in neuroscience}, vol. 10, pp. 196, 2016.

\bibitem{shah2018temple}
Vinit Shah, Eva Von~Weltin, Silvia Lopez, James~Riley McHugh, Lillian Veloso,
  Meysam Golmohammadi, Iyad Obeid, and Joseph Picone,
\newblock ``The temple university hospital seizure detection corpus,''
\newblock {\em Frontiers in neuroinformatics}, vol. 12, pp. 83, 2018.

\bibitem{velivckovic2017graph}
Petar Veli{\v{c}}kovi{\'c}, Guillem Cucurull, Arantxa Casanova, Adriana Romero,
  Pietro Lio, and Yoshua Bengio,
\newblock ``Graph attention networks,''
\newblock {\em arXiv preprint arXiv:1710.10903}, 2017.

\bibitem{finn2017model}
Chelsea Finn, Pieter Abbeel, and Sergey Levine,
\newblock ``Model-agnostic meta-learning for fast adaptation of deep
  networks,''
\newblock in {\em International conference on machine learning}. PMLR, 2017,
  pp. 1126--1135.

\bibitem{jasper1958ten}
Herbert~H Jasper,
\newblock ``The ten-twenty electrode system of the international federation,''
\newblock {\em Electroencephalogr. Clin. Neurophysiol.}, vol. 10, pp. 370--375,
  1958.

\bibitem{vaswani2017attention}
Ashish Vaswani, Noam Shazeer, Niki Parmar, and et~al.,
\newblock ``Attention is all you need,''
\newblock {\em Advances in neural information processing systems}, vol. 30,
  2017.

\bibitem{maas2013rectifier}
Andrew~L Maas, Awni~Y Hannun, Andrew~Y Ng, et~al.,
\newblock ``Rectifier nonlinearities improve neural network acoustic models,''
\newblock in {\em International conference on machine learning}. PMLR, 2013,
  vol.~30, p.~3.

\bibitem{rajendran2020meta}
Janarthanan Rajendran, Alexander Irpan, and Eric Jang,
\newblock ``Meta-learning requires meta-augmentation,''
\newblock {\em Advances in Neural Information Processing Systems}, vol. 33, pp.
  5705--5715, 2020.

\end{thebibliography}
\end{document}